\documentstyle[preprint,prb,aps]{revtex}
\tightenlines
\def\nb{{{\bar n}}}
\def\rr{{{\bf r}}}
\def\rrp{{{\bf r^\prime}}}
\def\nup{{{n_\uparrow}}}
\def\ndo{{{n_\downarrow}}}

\begin{document}

\title{ Finite-size effects and the stabilized spin-polarized jellium model
for metal clusters}

\author{M.~Payami }
\address{Center for Theoretical Physics and Mathematics,
Atomic Energy Organization of Iran, P.~O.~Box 11365-8486, Tehran, Iran}

\maketitle
\begin{abstract}

In the framework of spherical geometry for jellium and local spin density
approximation, we have obtained the equilibrium $r_s$ values, 
$\bar{r}_s(N,\zeta)$, of neutral and singly ionized ``generic" $N$-electron
clusters for their various spin polarizations, $\zeta$. Our results reveal
that $\bar{r}_s(N,\zeta)$ as a function of $\zeta$ behaves differently
 depending
on whether $N$ corresponds to a closed-shell or an open-shell cluster. That
is, for a closed-shell one, $\bar{r}_s(N,\zeta)$ is an increasing function of
$\zeta$ over the whole range $0\le\zeta\le 1$, and for an open-shell one, 
it has a decreasing part corresponding to the range $0<\zeta\le\zeta_0$, where 
 $\zeta_0$ is a polarization that the cluster assumes in a configuration
 consistent with Hund's first rule.
 In the context of the stabilized spin-polarized jellium model, our 
 calculations based on these equilibrium $r_s$ values, $\bar{r}_s(N,\zeta)$,
 show that instead of the maximum spin compensation (MSC) rule, Hund's first
 rule
 governs the minimum-energy configuration. We therefore conclude that the
 increasing behavior of the equilibrium $r_s$ values over the whole range of 
$\zeta$ is a necessary
condition for obtaining the MSC rule for the minimum-energy configuration;
and the only way to end up with an increasing behavior over the whole range of
$\zeta$ is to break the spherical geometry of the jellium background.
This is the reason why the results based on simple jellium with spheroidal
or ellipsoidal geometries show up MSC rule. 

\end{abstract}

\section{Introduction}
\label{sec1}
Since the production and study of sodium clusters by Knight {\it et al.},
\cite{knight84} the physics of metal clusters has attracted much interest.
\cite{deHeer93,Brack93}
Metal clusters are composed of atoms and have properties that are different
from both a single atom and the bulk metal. However, when increasing the
size of the cluster, its properties evolve to those of the bulk. The many-body
technique suitable for these systems is the density functional theory (DFT).
\cite{Kohn64,Sham65,Parr,Dreizler} It is a well-known fact that the properties
of alkali metals are
dominantly determined by the delocalized valence electrons. In these metals,
the pseudopotentials of the ions do not significantly affect the electronic
structure, because the Fermi wavelengths of the valence electrons become
much larger than the metal lattice constants if one replaces the ions with
a uniform positive charge background. This fact allows us to replace the
discrete ionic structure by a homogeneous positive charge background. This
approximation is known as the jellium model (JM). The simplest way of applying
the JM to metal clusters is to replace the ions of an $N$-atom cluster by
a sphere of uniform positive charge density and radius $R=(zN)^{1/3}r_s$,
where $z$ is the valence of the atom and $r_s$ is the bulk value of the
Wigner--Seitz radius of the metal. However, since the ionic density
near the surface of a metal differs from that of the bulk region, one may
resort to the diffuse jellium model (dif-JM) \cite{Rubio91} in which the
density of the jellium background
falls to zero, from the bulk value, within a length of a few atomic sizes.
Application of the dif-JM to metal clusters results in a better
agreement of the theory and experiment over the JM results.
However, in spite of its simplicity and success in predicting some properties
of bulk metals and metal clusters, the JM, which was originally developed for
bulk metals, has some drawbacks. \cite{Lang70,Ashlang67}
In order to overcome the deficiencies of the JM, one should take some details
of the ionic structure into account. Among the various methods of improvement,
the first attempts that kept the simplicity of the JM and overcame some
of the deficiencies of the JM resulted in the development of the stabilized
jellium model (SJM)\cite{Perdew90} or pseudojellium model.
 \cite{diazshore89,Rose91} The SJM was
applied to metal clusters \cite{Braj93,Perdew93} and improved
some  results
of the simple JM. Montag {\it et al.},\cite{Montag94} in their
structure-averaged
jellium model (SAJM), which added the ionic surface energy to the SJM
energy functional, made it more suitable for metal clusters.

In other approaches, some researchers relax the spherical geometry and use
the JM with spheroidal or ellipsoidal shapes, \cite{Ek91,Hir94,Yano95}
which are
suitable for open-shell clusters. Relaxing the shape of the jellium background
as well as its density distribution while keeping charge neutrality at every
point in space, called the ultimate jellium model (UJM), was introduced
by Koskinen {\it et al.}\cite{Kos95}
Since the bulk density in the UJM ($r_s=4.18$) is close to that of sodium,
its results can be compared with experiments on Na metal clusters.
\cite{kappes,homer}
However, since the jellium density and the electron density in the UJM are
locally equal everywhere, the UJM cannot describe the ionized clusters.

In a recent work, by taking the spin degrees of freedom into account in the
process of stabilization, we have generalized the SJM to the
stabilized spin-polarized jellium model (SSPJM). \cite{Pay98}
In the SJM, the equilibrium bulk density is a free parameter and the
experimental value is used for it. However, in the SSPJM the equilibrium
bulk density parameter, $\bar{r}_s(\infty,\zeta)$, is polarization dependent
and to the best of our knowledge, no experimental data are available for it.
Therefore in the SSPJM we take

\begin{equation}
\bar{r}_s^{\rm X}(\infty,\zeta)=\bar{r}_s^{\rm X}(\infty,0)+\Delta
r_s^{\rm EG}(\infty,\zeta),
\label{eq1}
\end{equation}
in which $\bar{r}_s^{\rm X}(\infty,0)$ is the equilibrium bulk value for the
spin-compensated system which takes the experimental value of metal X, and
 $\Delta r_s^{\rm EG}(\infty,\zeta)$ is obtained by the application
of the local spin-density approximation (LSDA) to the infinite electron-gas
 system. All equations throughout this paper are expressed in Rydberg atomic
 units. It turns out that $\Delta r_s^{\rm EG}(\infty,\zeta)$ is an
  increasing
 function of $\zeta$. By taking
$\zeta=(N_\uparrow-N_\downarrow)/N$ for a monovalent metal cluster, and
calculating the corresponding $\Delta r_s^{\rm EG}(\infty,\zeta)$ , it is
possible
to determine the appropriate radius for the spherical jellium through
$R(N,\zeta)=N^{1/3}\bar{r}_s(\infty,\zeta)$. Therefore, any variation
in $N_\uparrow$ and $N_\downarrow$, keeping $N=N_\uparrow+N_\downarrow$
constant, leads to different values for $\zeta$ and thereby, different
values for $R(N,\zeta)$.

Application of this model for the energy calculation of metal clusters
show that the energy is minimized for a configuration with
maximum spin-compensation (MSC). \cite{Pay98}
That is, for clusters with an even number of electrons, $N$, the number of
 up-spin electrons, $N_\uparrow$, equals the number of down-spin electrons,
 $N_\downarrow$; and for an odd number of electrons,
 $N_\uparrow - N_\downarrow =1$. This MSC rule leads to the fine structure
 in $\Delta_2(N)$ (see Fig. 5 of Ref. \protect\ref{Pay98}) and the odd--even
alternations  in the ionization energies [Fig. 7(c) of Ref. \protect\ref{Pay98}].

 In the present paper, keeping the spherical geometry for the jellium, we
 have investigated the finite-size effects on the equilibrium $r_s$ values,
 and their consequences on the SSPJM calculations.
 First, using the LSDA, we have found $\bar{r}_s(N,\zeta)$, the equilibrium 
$r_s$ values of
 the closed-shell $N$-electron neutral and singly ionized ``generic" clusters
 with $N=2,8,18,20,34,40$ for all possible polarizations.
By generic we mean no specific metal but a simple jellium sphere which can
 assume its equilibrium size for any given value of $N$. 
 By fitting
 the calculated results to a polynomial with even powers of $\zeta$ (in the
  absence of magnetic fields, the physical properties are invariant under the
 transformation $\zeta \rightarrow -\zeta$), we have found two analytic
 equations for $\Delta r_s(\zeta)$, one for neutral and the other for singly
 ionized clusters. Employing these analytic equations in the
  SSPJM  calculations for Na clusters shows that as before, the MSC rule is
 at work and
the results do not show any significant changes over our previous results
on the ionization energies. \cite{Pay98}
 In the next step, we have found the values $\bar{r}_s(N,\zeta)$ for all
 different
 neutral and singly ionized generic clusters
 ($N\le42$) with different spin configurations ($0\le\zeta\le 1$), and
 have shown that $\bar{r}_s(N,\zeta)$ behaves differently for open-shell
  and closed-shell clusters. That is, we have found that for a closed-shell
 cluster (and its two nearest neighbors) it is an increasing function of
  $\zeta$ over the whole interval $0\le\zeta\le 1$, whereas for an open-shell
 cluster (except for the two nearest neighbors of a closed-shell cluster), it
  has a decreasing behavior over $0\le\zeta\le\zeta_0$, and
 an increasing behavior over $\zeta_0\le\zeta\le 1$.
 However, in both
 open-shell and closed-shell clusters, the global minimum of energy, i.e.,
 the ground-state energy,
 corresponds to a configuration in which $\bar{r}_s(N,\zeta)$
 is a minimum. Here, $\zeta_0$ corresponds to an electronic
 configuration for which Hund's first rule is satisfied.
 By subtracting the ground-state energy of a singly ionized generic cluster
 from that of its neutral counterpart, we obtain the ionization energy of that
 generic
 cluster.  Calculation of these ionization energies for 
  $N\le 42$ shows a good agreement with experimental results on Na clusters.
 We see that, in the ionization-energy plot, although the saw-toothed behavior
 remains, the pronounced shell effects near closed shells, seen in the simple
 JM results, are substantially reduced.
 Thanks to the appreciable reduction of the shell effects in the
 ionization energy results of the above-mentioned
 calculations, we have performed the SSPJM calculations using the set of
 values $\bar{r}_s(N,\zeta)$. The
 results show that, instead of the MSC rule, Hund's first rule is governing the
 ground-state configuration.
 We have also performed simple JM-LSDA calculations for the ground-state
 energies of Na clusters but, instead of using the
 ordinary bulk $r_s$ value (3.99), we have used the increasing function
 $\bar{r}_s(\infty,\zeta)$ as
 in Eq. (\ref{eq1}). The results show that, here also, Hund's first rule
 remains at work
 and therefore, we conclude that with spherical geometries, the increasing
 behavior of
 $\Delta r_s^{\rm EG}(\infty,\zeta)$ is 
a necessary condition (but not sufficient) for realizing the MSC rule, and
the energy corrections over the simple JM due to stabilization are also needed.  
Therefore, in order to improve the SSPJM results one should insist on
the increasing behavior for $\bar{r}_s(N,\zeta)$. The only way which 
guarantees such behavior for all values of $N$ is to relax 
the spherical constraint on the geometry
 and consider ellipsoidal shapes for open shells. Using the LSDA
 and ellipsoidal geometry, one could obtain $\bar{r}_s(N,\zeta)$, for a
 given set of values of $N$ and $\zeta$, by finding the values of the
  ellipsoid
 axes which correspond to the minimum of energy. We expect that in this
 case $\bar{r}_s(N,\zeta)$ will be an increasing function of $\zeta$ for
 both open-shell and closed-shell clusters. Then, if one performs the SSPJM
 calculations for ellipsoidal clusters using these new increasing functions
 $\bar{r}_s(N,\zeta)$,
 one would obtain the MSC rule (and thereby the odd--even
 alternations) with improved results.

 The organization of this paper is as follows. Section \ref{sec2} is devoted
 to calculational schemes. In Sec. \ref{sec3} we present the results of our
 calculations. In Sec. \ref{sec4} we conclude the work.

 \section{calculational Schemes}
 \label{sec2}

The energy functional in the SSPJM is given by \cite{Pay98}

\begin{eqnarray}
E_{\rm SSPJM}[\nup,\ndo,n_+]&=&
E_{\rm JM}[\nup,\ndo,n_+]+(\varepsilon_{\rm M}(\nb)+\bar w_R(\nb))
\int d\rr\;n_+(\rr) \nonumber \\
  &&+\langle\delta v\rangle_{\rm WS}(\nb)\int
d\rr\;\Theta(\rr)[n(\rr)-n_+( \rr)],
\label{eq2}
\end{eqnarray}
in which $E_{\rm JM}$ is the energy fuctional of the simple JM,
$\varepsilon_{\rm M}$ is the Madelung energy, $\bar{w}_R$ is the average
value of the repulsive part of the pseudopotential, and $\langle \delta v
\rangle_{\rm WS}$ is the average of the difference potential over the
  Wigner--Seitz cell and the difference potential, $\delta v$, is defined
as the difference between the pseudopotential of a lattice of ions and
the electrostatic potential of the jellium background. $\nup(\rr)$ and
 $\ndo(\rr)$
are, respectively, the up-spin and down-spin electron densities with the total
electron density given by $n(\rr)=\nup(\rr) + \ndo(\rr)$; and $\bar{n}$ is
the uniform jellium density which is $\zeta$-dependent; for our previous
SSPJM calculations we have used the values given by Eq. (\ref{eq1}).
In the spherical JM, the functions will depend only on the
radial variable. The function $\Theta(\rr)$ becomes a simple
 radial  step function, $\theta(R-r)$, where $R=N^{1/3}\bar{r}_s$ with
 $\bar{r}_s=(3/4\pi\bar{n})^{1/3}$. In the case of jellium with a
 sharp boundary, $n_+(r)=\bar{n}\theta(R-r)$, whereas for the diffuse case
 we take \cite{Rubio91}

\begin{equation}
n_+(r)=\left\{\begin{array}{l}
              \nb\{1-(R+t)e^{-R/t}[\sinh(r/t)]/r\},\;\;\;r\le R\\
              \nb\{1-((R+t)/2R)(1-e^{-2R/t})\}R e^{(R-r)/t}/r,\;\;\;r>R,
\end{array}
        \right.
\label{eq3}
\end{equation}
 where $t$ is a parameter related to the surface thickness.

 To evaluate the total energy of a cluster, we solve the Kohn--Sham (KS)
 equations self-consistently. The effective potential in the KS
 equations for the SSPJM calculations is given by

 \begin{equation}
 v_{\rm eff}^\sigma([\nup,\ndo,n_+];\rr)=\phi([n,n_+];\rr)+
 v_{xc}^\sigma([\nup,\ndo];\rr)+\langle\delta v\rangle_{\rm WS}(\bar{n})
 \Theta(\rr),
\label{eq4}
\end{equation}
where

\begin{equation}
\phi([n,n_+];\rr)=2\int d\rrp\;\frac{[n(\rrp)-n_+(\rrp)]}{\mid\rr-\rrp\mid}.
\label{eq5}
\end{equation}
which appears in the electrostatic part of the total energy of the simple JM
energy functional

\begin{eqnarray}
E_{\rm JM}[\nup,\ndo,n_+]&=&T_s[\nup,\ndo]+E_{xc}[\nup,\ndo] \nonumber\\
&&+\frac{1}{2}\int d\rr\;\phi([n,n_+];\rr)[n(\rr)-n_+(\rr)],
\label{eq6}
\end{eqnarray}
and

\begin{equation}
v_{xc}^\sigma([\nup,\ndo];\rr)=\frac{\delta E_{xc}}{\delta n_\sigma(\rr)},
\;\;\;\;\;\;\sigma=\uparrow,\downarrow.
\label{eq7}
\end{equation}

For $E_{xc}$ we use the LSDA with the Perdew--Wang parametrization for the
correlation part.\cite{Perwang92}

\section{Results and Discussions}
\label{sec3}

In the first step, by solving the KS equations for spherical geometries of
the jellium in the LSDA, and finding the
minimum energies, we have obtained  $\bar{r}_s(N,\zeta)$, the equilibrium
 $r_s$ values of the closed-shell neutral and singly ionized jellium clusters
 with $N=2,8,18,20,34,40$ electrons. In the
 calculations we have considered all different possible polarizations.
 That is, for an
 $N$-electron cluster ($N=N_\uparrow+N_\downarrow$), we have found
 $\bar{r}_s(N,\zeta)$ for all different polarizations $(0\le\zeta\le 1)$
 corresponding to the
 configurations $N_\downarrow=N_\uparrow$, $N_\downarrow=N_\uparrow-1$,
 $N_\downarrow=N_\uparrow-2$, \ldots, $N_\downarrow=0$.
 The results are shown in Figs. \ref{fig1}(a) and \ref{fig1}(b).
  We see that for these closed-shell
clusters, $\bar{r}_s$ is an increasing function of $\zeta$.
By a least-square
 fitting of the results to the polynomial

\begin{equation}
r_s(\zeta)=a_0 + a_2\zeta^2 + a_4\zeta^4 + a_6\zeta^6,
\label{eq9}
\end{equation}
we have obtained $a_0=4.28$, $a_2=2.15$, $a_4=-2.41$, $a_6=1.84$ for neutral,
and $a_0=4.62$, $a_2=1.49$, $a_4=-3.31$, $a_6=3.92$ for singly ionized
clusters.
In Figs. \ref{fig1}(a) and \ref{fig1}(b), we have also shown plots using
 Eq. (\ref{eq9})
for the two sets of the coefficients and compared  with
the bulk function $\bar{r}_s^{\rm EG}(\infty,\zeta)$.
The values of $a_0$ show that the simple JM with LSDA
predicts a larger atomic
spacing for a cluster than the bulk value ($a_0>4.18$).
Regardless of the $a_0$ values
(since we are studying the variations of $r_s$ with respect to $\zeta$),
we have performed our SSPJM calculations using

\begin{equation}
\Delta r_s(\zeta)=a_2 \zeta^2 + a_4 \zeta^4 + a_6 \zeta^6,
\label{eq10}
\end{equation}
with corresponding coefficients for neutral and singly ionized clusters.
Figure \ref{fig2} compares the plots of Eq. (\ref{eq10}) for neutral and
singly ionized clusters with $\Delta r_s^{\rm EG}(\infty,\zeta)$ of the bulk
 jellium.
 We see that they are increasing functions of $\zeta$.
Using Eq. (\ref{eq10}), we have performed the dif-SSPJM calculations for Na
clusters with $\bar{r}_s(\infty,0)=3.99$. After the self-consistent
calculations of the KS equations, we have obtained the dif-SSPJM 
ground-state total
 energies of the neutral and singly ionized Na clusters ($N\le 42$).
The calculations show that the MSC rule is governing the ground-state
configuration.
In the dif-SSPJM calculations, we have taken $t=1$. In Fig. \ref{fig3}
we have plotted the ionization energies and compared them with our previous
results\cite{Pay98} and also with the experimental values.
As is shown, there are no significant changes over the previous results.

In the second step, using JM-LSDA we have found the values
$\bar{r}_s(N,\zeta)$ for all different neutral and singly ionized jellium
clusters ($N\le 42$) with different spin polarizations. Here, for a cluster
with specified values of $N$ and $\zeta$, the value $\bar{r}_s(N,\zeta)$
minimizes the total energy of the cluster.
In Fig. \ref{fig4} we have shown the values of $\bar{r}_s(N,\zeta)$
and corresponding total energies per electron, $\bar{E}(N,\zeta)/N$, as
functions of the number of the electrons, $N$, for neutral
jellium clusters. The two plots show the same structure.
That is, the maxima and minima of these two plots
correspond to the same values of $N$.
The values of the ground-state energies
correspond to polarizations $\zeta_0$ consistent with Hund's first rule and
form the
Hund curve.
In each of the plots, the uppermost value for a given $N$ corresponds to the
configuration with MSC. The values in between correspond to intermediate
polarizations. We therefore conclude that for an open-shell cluster the
 functions $\bar{r}_s(N,\zeta)$ and $\bar{E}(N,\zeta)/N$
  have minima for a polarization $\zeta_0 \ne 0$ which
is consistent with Hund's first rule.
\cite{slater,tate74,tate94,toshi95} That is,
for open-shell clusters (except for the nearest neighbors to the closed-shell
 cluster)
$\bar{r}_s$ and $\bar{E}(N,\zeta)$ are  decreasing functions of $\zeta$ for
 $0\le\zeta\le\zeta_0$ and
increasing functions for $\zeta_0\le\zeta\le 1$. In the bulk jellium,
the function $\bar{r}_s^{\rm EG}(\infty,\zeta)$ is an increasing function
 over the whole range $0\le\zeta\le 1$.
In Fig. \ref{fig5} we have compared the functions $\bar{r}_s(27,\zeta)$
and $\bar{r}_s(34,\zeta)$ with $\bar{r}_s^{\rm EG}(\infty,\zeta)$. We see
that for the closed-shell cluster
 ($N=34$) the volume is an increasing
function of $\zeta$ on the whole range $0\le\zeta\le 1$. But for
the open-shell cluster ($N=27$) the volume decreases with increasing $\zeta$
in the range $0<\zeta\le 7/27$ and expands with $\zeta$ in the range
$7/27\le\zeta\le 1$. The value $\zeta_0=7/27$ is due to the half-filled shell
($l=3$) of the up-spin band.
 Also, we note that
 $\bar{r}_s(27,\zeta) > \bar{r}_s^{\rm EG}(\infty,\zeta)$
over the whole range $0\le\zeta\le 1$.
Perdew {\it et al.} have also studied the equilibrium size of 
spherical clusters of stabilized jellium from a different point of view.
\cite{Perdew93}

Figure \ref{fig6} shows the ground state $\bar{E}/N$ and its corresponding
$\bar{r}_s$ for
 neutral and singly ionized jellium clusters as functions of $N$, the number
 of electrons.
The energies correspond to the equilibrium $r_s$ values
 for configurations consistent with Hund's first rule.  Here, $N$
 is the number of electrons which for neutral clusters equals the number
 of the ions, and for singly ionized clusters it is one less than the number
 of the ions. The ionization energy of an $N$-atom
  jellium cluster is given by

\begin{equation}
I(N)=\bar{E}^{\rm ion}(N-1,\zeta^\prime_0)-
     \bar{E}^{\rm neut}(N,\zeta_0),
\label{eq11}
\end{equation}
where $\bar{E}^{\rm neut}(N,\zeta_0)$ is the ground-state energy of the
neutral $N$-electron jellium cluster which occurs at polarization $\zeta_0$
consistent with Hund's rule, and $\bar{E}^{\rm ion}(N-1,\zeta^\prime_0)$
is the ground state energy of the singly ionized cluster with $N-1$ electrons
and $N$ positive ions. Obviously, the polarizations $\zeta_0$ and
 $\zeta^\prime_0$ are not the same. In Fig. \ref{fig7} we have plotted
 the ionization energies of the jellium clusters ($N\le 42$) obtained from
 Eq. (\ref{eq11}), and compared them with the results obtained from the simple JM
 calculations (with $r_s=3.99$) and experiment on Na clusters. The
 relevance of this
 comparison with Na results is due to the fact that the equilibrium $r_s$
 value of the bulk jellium, 4.18, is very close to that of the bulk Na.
 Except for very small clusters, the results based on the equilibrium $r_s$
  values show a better agreement
 with experiment than the results of simple JM calculations for Na clusters.

 We saw, in the second step, that if the jellium clusters assume their
 individual equilibrium volumes, then the ionization energies improve.
 Therefore, we are naturally led to define a new set of $\Delta r_s$ for
our SSPJM calculations which are obtained from

\begin{equation}
\Delta r_s(N,\zeta)=\bar{r}_s(N,\zeta)-a_0.
\label{eq12}
\end{equation}
Here, $a_0=4.28$ for neutral and $a_0=4.62$ for singly ionized clusters;
$\bar{r}_s(N,\zeta)$ is the equilibrium $r_s$ value for a cluster with given
values of $N$ and $\zeta$. Using Eq. (\ref{eq12}), we have performed the SSPJM
calculations for Na. The results of calculations show that here (in contrast to
the MSC rule) the energy of a cluster is minimized for a configuration
consistent with Hund's first rule. This argument is valid for both neutral
and singly ionized clusters. This behavior has its roots in the decreasing
behavior of $\bar{r}_s(N,\zeta)$ over the range $0<\zeta\le\zeta_0$ for
 open-shell clusters.
We have also checked whether or not taking account of the volume change as a function
of polarization, as given by Eq. (\ref{eq1}), in the simple JM calculations
gives rise to the MSC rule.
The results show that Hund's first rule remains at work and therefore we
conclude that, in order to obtain MSC configuration for the ground state,
not only $\Delta r_s$ should be an increasing function of $\zeta$ but one
 should also include the two corrections, due to the stabilization, in the simple JM
energy. That is, one should use the SSPJM energy along with an increasing
function for $\Delta r_s$. This condition will be met if one relaxes
the spherical constraint on the jellium and considers ellipsoidal shapes for
open-shell clusters. It is then possible to obtain $\bar{r}_s(N,\zeta)$ for
a given set of values of $N$ and $\zeta$, by finding the values of the
ellipsoid axes that minimize the total energy. We expect that under these
conditions $\bar{r}_s(N,\zeta)$ becomes an increasing function of $\zeta$
for all clusters. This expectation is due to the fact that addition of a
nonsymmetric perturbation to a spherical potential in a single-particle
Hamiltonian lifts the orbital degeneracies and each degenerate level splits
into ($2l+1$) new levels with different energies. Each of these levels will
contain at most two electrons with opposite spins. Then, any increment in the
polarization is accompanied by a transition of a spin-down electron to an
unoccupied level and a successive spin-flip. This is the only way that one
can increase the polarization consistent with Pauli's exclusion principle.
This process resembles the process of increasing the polarization in a
closed-shell spherical cluster which results in increasing its equilibrium
$r_s$ value. It is a well-known fact that the open-shell clusters lose their
spherical geometry due to the Jahn--Teller effect,\cite{jahn} and that is why
the MSC rule and the odd--even alternations are observed in experimental data
 for alkali metal clusters.
In short, using a new set of increasing functions
$\bar{r}_s(N,\zeta)$ -- which is obtained using ellipsoidal geometries
for simple JM and LSDA-- in the SSPJM calculations for ellipsoidal geometries
of the jellium will lead to the MSC configuration for the ground state of
the cluster. Work in this direction is in progress.

Finally, it should be mentioned that in the context of the SSPJM one could
fix, at the beginning, the pseudopotential core radius for a bulk system of
a given species with a given fixed polarization; and then use this value in the
 SSPJM energy functional for a finite cluster (transferability condition
on the pseudopotential).\cite{Perdew93} 
Then, one could obtain the equilibrium radius of
the jellium by finding that $r_s$ value which minimizes the total energy.
We have performed such calculations and obtained further agreement to the
experimental results, which will appear elsewhere.    
\section{Conclusion}
\label{sec4}

In this paper, keeping the spherical geometry, we have considered the
finite-size effects on the equilibrium $r_s$ values. Our calculations show
that for a given $N$-electron cluster, the quantity $\bar{r}_s(N,\zeta)$
behaves differently for an open-shell and a closed-shell cluster. That is,
this equilibrium $r_s$ value is an increasing function of $\zeta$ over the
whole range $0\le\zeta\le 1$ for a closed-shell cluster; whereas for an
open-shell cluster it is a decreasing function over $0<\zeta\le\zeta_0$
and an increasing function over $\zeta_0\le\zeta\le 1$. Here, $\zeta_0$
is a polarization corresponding to a configuration consistent with Hund's
first rule. Our SSPJM calculations based on equilibrium $r_s$ values
show that, in contrast to the MSC rule, Hund's first rule is at work.
This behavior is due to the fact that $\bar{r}_s(N,\zeta)$ has a decreasing
part for an open-shell cluster.
We therefore conclude that to realize the MSC rule in the SSPJM calculations
with $\bar{r}_s(N,\zeta)$, and thereby the
odd--even alternation, one should lift the spherical constraint on the jellium
and let it assume ellipsoidal shapes.
\acknowledgements{
The author would like to thank John P. Perdew for reading  
the manuscript and his helpful discussions on the subject. He also thanks N.
 Nafari for his useful discussions.}

\newpage
\begin{figure}
\caption{The equilibrium $r_s$ values in atomic units as functions of the
polarization $\zeta$. Large squares, diamonds, crosses, pluses, small squares,
triangles correspond to clusters with $N$=2, 8, 18, 20, 34, 40 electrons,
respectively, for (a) neutral, and (b) singly ionized closed-shell clusters.
The solid lines and the dashed lines correspond to fitted and bulk values,
respectively.}
\label{fig1}
\end{figure}

\begin{figure}
\caption{The increments of $r_s$ in atomic units as functions of $\zeta$.
The dotted and dashed lines correspond to the fitted result [ see Eq.
(\protect\ref{eq10}) ] for neutral and singly ionized clusters, respectively.
The solid line corresponds to the bulk jellium.}
\label{fig2}
\end{figure}

\begin{figure}
\caption{The ionization energies as functions of the number of atoms for Na
clusters. The diamonds correspond to the dif-SSPJM results obtained using Eq.
(\protect\ref{eq1}) and the triangles correspond to the dif-SSPJM results
obtained using Eq. (\protect\ref{eq10}). The small squares are experimental
results.}
\label{fig3}
\end{figure}

\begin{figure}
\caption{The equilibrium $r_s$ values in atomic units and their corresponding
total energies per electron in electron volts as functions of the number of
electrons for neutral clusters with different spin configurations.
 The triangles correspond to the
configurations consistent with Hund's first rule, and the uppermost symbols
for each $N$ correspond to the MSC configurations. The symbols between the
Hund and the MSC configurations correspond to intermediate polarizations.
The horizontal dashed line below the $\bar{r}_s$ curve corresponds to the
value $r_s=4.18$ of the bulk jellium.  }
\label{fig4}
\end{figure}

\begin{figure}
\caption{The equilibrium $r_s$ values in atomic units as functions of $\zeta$.
The solid squares and diamonds correspond to the closed-shell ($N=34$) and
open-shell ($N=27$) clusters, respectively. The dashed line corresponds to the
bulk which is obtained using Eq. (17) of Ref. \protect\ref{Pay98}.
As is seen, the equilibrium $r_s$ of the open-shell cluster has a decreasing
part before $\zeta_0=7/27$, whereas for the closed-shell cluster it is
an increasing function over the whole range $0\le\zeta\le 1$.    }
\label{fig5}
\end{figure}

\begin{figure}
\caption{The equilibrium $r_s$ values in atomic units and their corresponding
total energies per electron in electron volts for the ground-state
 configurations (i.e., the configurations consistent with Hund's first rule)
 of $N$-electron clusters. The solid squares correspond to neutral and the
 diamonds correspond to singly ionized $N$-electron clusters. As is expected,
 for small clusters the differences are high and for large clusters these
 differences approach zero.      }
\label{fig6}
\end{figure}

\begin{figure}
\caption{The ionization energies in electron volts. The diamonds correspond
to the simple JM results for Na using the bulk value $r_s=3.99$.
The small solid squares correspond to the ionization energies of jellium
 clusters in their equilibrium states obtained using Eq. (\protect\ref{eq11}).
It shows a good agreement with the experimental values of Na, which are shown
as the large squares. }
\label{fig7}
\end{figure}

\end{document}